\documentclass[12pt]{article}

\usepackage{graphicx}
\usepackage{algorithm}
\usepackage{algorithmic}
\usepackage{xspace}
\usepackage{subfigure}
\usepackage{url}
\newtheorem{definition}{Definition}

\long\def\comment#1{}
\newcommand{\stitle}[1]{\vspace{1ex} \noindent{\bf #1}}
\newcommand{\nop}[1]{}

\begin{document}

\title{Detecting Priming News Events}
% \numberofauthors{1}
\author{Di Wu, Yiping Ke, Jeffrey Xu Yu, Zheng Liu \\
       The Chinese University of Hong Kong, Hong Kong, China \\
       \{dwu,ypke,yu,zliu\}@se.cuhk.edu.hk
}
\date{}

% \title{Detecting Priming News Events}
% \author{Submitted for Blind Review}

\maketitle

\begin{abstract}

We study a problem of detecting priming events based on a time series
index and an evolving document stream. We define a priming event as an
event which triggers abnormal movements of the time series index,
i.e., the Iraq war with respect to the president approval index of
President Bush. Existing solutions either focus on organizing coherent
keywords from a document stream into events or identifying correlated
movements between keyword frequency trajectories and the time series
index. In this paper, we tackle the problem in two major steps. (1) We
identify the elements that form a priming event.  The element
identified is called influential topic which consists of a set of
coherent keywords. And we extract them by looking at the correlation
between keyword trajectories and the interested time series index at a
global level.  (2) We extract priming events by detecting and
organizing the bursty influential topics at a micro level. We evaluate
our algorithms on a real-world dataset and the result confirms that
our method is able to discover the priming events effectively.
\end{abstract}

\section{Introduction}

\label{sec:introduction}

With the increasing text information published on the news media
and social network websites, there are lots of events emerging
every day. Among these various events, there are only a few that
are priming and can make significant changes to the world. In this
paper, we measure the priming power of the events based on some
popular index that people are usually concerned with.

\stitle{Governor's Approval Index } Every activity of the governor can
generate reports in the news media and discussions on the web.
However only a few of them will change people's attitude towards this
governor and pose impacts on his/her popularity, i.e., approval index
\cite{CambridgeJournals:5827084}. Since these approval indices are the
measures of the satisfaction of citizens and crucial for the
government, people would be more interested in knowing the events that
highly affect the approval index than those with little impact. 

\stitle{Financial Market Index} There are a lot of events related to
the company or the stock market happening while only some of them
would change the valuation of the company in investors' mind. The
greedy/fear will drive them to buy/sell the stock and eventually
change the stock index. Therefore, the investors will be eager to
know and monitor these events as they happen.

These real world cases indicate that there is a need to find out
the priming events which drive an interested time series index.
Such priming events are able to help people understand what is
going on in the world. However, the discovery of such priming
events poses great challenges: 1) with the tremendous number of
news articles over time, how could we identify and organize the
events related to a time series index; 2) several related events
may emerge together at the same time, how can we distinguish their
impact and discover the priming ones; 3) as time goes by, how
could we track the life cycle of the priming events, as well as
their impact on the evolving time series index.
\begin{figure}
\centering
\hspace*{-0.8cm}
\includegraphics[width=8cm,height=4cm]{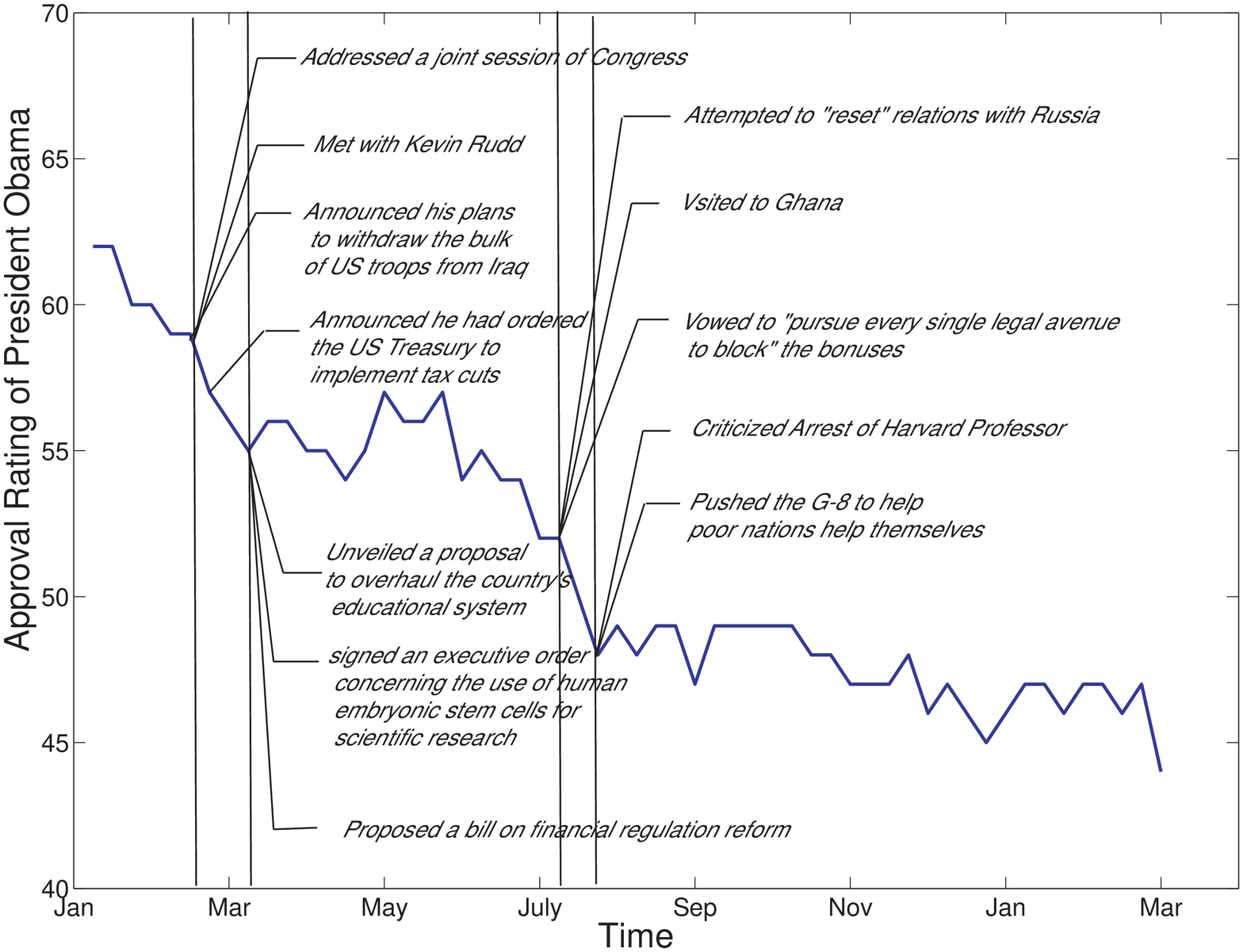}
\label{fig:obama_index} \caption{Approval Index of President
Obama}
\hspace*{-0.6cm}

\end{figure}
%%%%

Some existing work has focused on discovering a set of topics/events
from the text corpus \cite{topic.detection,DBLP:conf/vldb/FungYYL05}
and tracking their life cycle \cite{DBLP:conf/kdd/LeskovecBK09}. But
these methods make no effort to guarantee these events are influential
and related to the index that people are concerned with. There is
another stream of work considering the relationship between the
keyword trajectory and interested time series
\cite{DBLP:conf/kdd/GruhlGKNT05,DBLP:conf/wise/WuFYL08,stockriskrank}.
However, these work can only identify a list of influential keywords
for users and do not consider to organize these words into some high
level meaningful temporal patterns (i.e., events).

In this paper, we study the problem of detecting priming events from a
time series index and an evolving document stream. In Figure
\ref{fig:obama_index}, we take the weekly approval index of US
President Obama from Jan 20, 2009 to Feb 28, 2010 as an example to
illustrate the difficulty of this problem. In Figure
\ref{fig:obama_index}, the approval index (blue line) evolves and
drops from 67\% to 45\% in the last 56 weeks. Particularly, in July,
2009, the index dropped from 54\% to 48\%. For a user who is
interested in politics and wants to know what event trigger this
significant change, he/she may issue a query ``President Obama'' to
the search engine. But the result will only be a list of news articles
indicating the events that President Obama participates in during
these periods. In Figure \ref{fig:obama_index}, we tag a small part of
them on the index. As we can see, in July, 2009, there are 5 events
including his attempt to reset relationship with Russia, help pool
nations, visit Ghana etc. Only with this information, we cannot
fulfill the user's need since we cannot differentiate the role that
each event plays to change the approval index in that time
period. This urges us to think about the following questions.  What
makes an event priming? Does it contain some elements which will
attract public eyes and change their mind? Besides, if such elements
do exist, could we find their existence evidence from other time
period and use them to justify the importance of the local event
containing them?

In this paper, we call such elements influential topics and use
them as basic units to form a priming event. Specifically, we
identify the influential topics at a global level by integrating
information from both a text stream and an index stream. Then at a
micro level, we detect such evidences and organize them into
several topic clusters to represent different events going on at
each period. After that, we further connect similar topic clusters
in consecutive time periods to form the priming events. Finally,
we rank these priming events to identify their influence to the
index.

The contributions of this paper are highlighted as follows.
\begin{itemize}
\item To the best of our knowledge, we are the first to formulate
the problem of detecting priming events from a text stream and a
time series index. A priming event is an object which consists of
three components: (1) Two timestamps to denote the beginning and
ending of the event; (2) A sequence of local influential topic
groups identified from the text stream; (3) a score representing
its influence on the index.

\item We design an algorithm that first discovers the influential
topics at a global level and then drills down to local time
periods to detect and organize the priming events based on the
influential topics.

\item We evaluate the algorithm on a real world dataset and the
result shows that our method can discover the priming events
effectively.
\end{itemize}

% what we do

The rest paper is organized as follows. Section \ref{sec:overview}
formulates the problem and gives an overview to our approach.
Section \ref{sec:dataTransform} discusses how we detect bursty
text features and measure the change of time series index. Section
\ref{sec:topicDetection} describes the influential topic detection
algorithm and Section \ref{sec:eventExtraction} discusses how we
use the influential topics to detect and organize priming events.
Section \ref{sec:experiment} presents our experimental result.
Section \ref{sec:relatedWork} reviews some related work and
Section \ref{sec:conclusion} concludes this paper.

\section{Problem Formulation}

\label{sec:overview}

\comment{
\begin{figure}
\centering
\includegraphics[width=8cm,height=5cm]{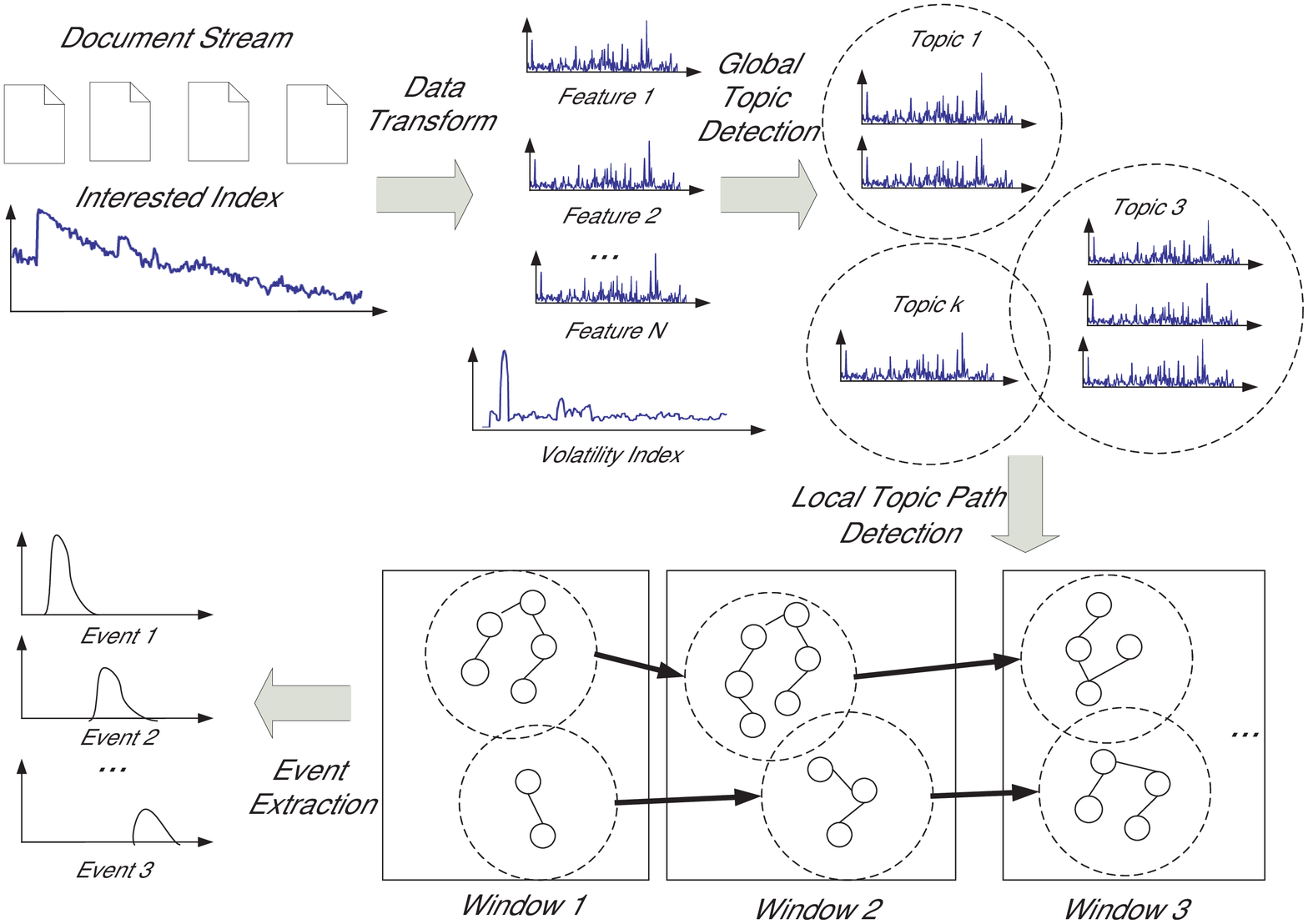}
\caption{Framework}
\label{fig:framework}
\end{figure}
}
% keyword frequency time series (FT)

% keyword bursty time series (BT)

% volatility index (VI)

% priming topics (PT)

% daily priming event (DPE)

% priming event path (PEP)

\comment{
\begin{table}[t]
\centering
\begin{tabular}{|ll|}
\hline {\bf Symbol} & {\bf Description}\\\hline
$\cal D$     &a text corpus\\
$d_i$     &a document, $d_i \in \cal D$\\
$t_i$     &the timestamp of $d_i$\\
$I$     &an interested index\\
$VI$     &the volatility index from $I$\\
$W$     &a set of windows\\
$w_i$     &a window, $w_i \in W$\\
$L$     &number of windows in $I$\\
$D_w$     &a set of documents in $w$, $D_w \in \cal D$\\
$F$     &a set of all different features\\
$f$     &a feature, $f \in F$\\
$F_w$     &a set of features in $w$, $F_w \in F$\\
$\theta$     &a set of influential topics\\
$C_w$     &feature clusters in $w$\\
$c_w$     &a feature cluster, $c_w \in C_w$\\
$PE$     &a set of priming events\\
$pe$     &a priming event, $pe \in PE$\\
\hline
\end{tabular}
\par
\caption{Notations} \label{table:term}
\end{table}
Table 1 summarizes the notations that would be used throughout this paper.
}

Let ${\cal D}=\{d_1, d_2,... \}$ be a text corpus, where each
document $d_i$ is associated with a timestamp. Let $I$ be the
interested index, which consists of $|W|$ consecutive and
non-overlapping time windows $W$. $\cal D$ is then partitioned
into $|W|$ sets according to the timestamp of the documents. Let $F$
be a set containing all different features in $\cal D$, where a
feature $f \in F$ is a word in the text corpus.

%%In the following section, we will first describe how we formulate
%%the problem and why it is formulated in such a way by presenting
%%some real life examples from the political domain.

% global topic detection

% why not detect events using vldb'05 and sigir'07

% why index information is important, define influential topic

Given the interested index $I$ and the text corpus $\cal D$, our
target is to detect the priming events that trigger the movement
of the index. As discussed in Section \ref{sec:introduction}, the
first step is to discover the influential topics. A possible
approach \cite{DBLP:conf/vldb/FungYYL05,DBLP:conf/sigir/HeCL07} is
to first retrieve all the topics from the documents using the
traditional approach in topic detection and tracking (TDT). Then
we detect the influential topics by comparing the strength of the
topic with the change of the index over time. We argue that this
approach is inappropriate because the topic detection is purely
based on the occurrence of the words and ignores the behaviors of
the index. Consider the feature \emph{worker}. Typical TDT
approach would consider its co-occurrences with other features
when deciding to form a topic. By enumerating all the
possibilities, it will form a topic with features such as
\emph{union} because \emph{worker union} frequently appears in
news documents. However, if we take the presidential approval
index into consideration, the most important event about
\emph{worker} related to the index would be that President Bush
increased the Federal Minimum Wage rate for workers ever since
1997. This event helped him to stop the continuous drop trend of
the approval index and make it stay above 30\%. Therefore, it is
more favorable to group \emph{worker} with \emph{wage} rather than
with \emph{union}. This example urges us to consider how to
leverage the information from the index to help us organize the
features into influential topics. These influential topics take in
not only the feature occurrence information but also the changing
behavior of the index. We formally define such topics as follows.

\begin{definition}
(\textbf{Influential Topics}) An influential topic $\theta_i$ is
represented by a set of  semantically coherent features
$F_{\theta_i} \subseteq F$ with a score indicating its influence
on the time series index $I$.
\end{definition}

% why not the topic is not enough, local events detection

% define local influential cluster

% why path detection and ranking

% define priming events

%In this paper, we design an algorithm which integrates the index information into influential topic detection and the implementation detail will be presented in section \ref{sec:topicDetection}.

%Without the loss of generality, assume that we have extracted all
%of the influential topics $\theta$ from the text corpus, a
%question rises immediately: how could we use these topics to
%represent the priming events.
% why one topic is not enough to represent an event
% correlated topic
% evolution

Based on the definition of influential topics, the next step is to
represent priming events using these topics. One simple and direct
way is to take each occurrence of a topic $\theta_i$ as a priming
event. However, this approach has one major problem. We observe
multiple topics at a time window $w$ and these topics are actually
not independent but correlated and represent the same on-going
event. Our topic detection algorithm may not merge them into a
single topic because they only co-occur at that certain window but
separate in other windows. For example, the topic of \{strike
target\} would appear together with the topic \{force troop
afghanistan\} in 2001. But in 2003, when the Iraq war starts, it
co-occurs with the topic \{gulf war\} instead. Therefore, in order
to capture such merge-and-separate behavior of topics, we define
the local topic cluster as follows.

\begin{definition}
(\textbf{Local Topic Cluster}) A local topic cluster $c_{w,i}$ in
$w$ consists of a set of topics which occur in highly overlapped
documents to represent the same event.
\end{definition}

%Assume we have identified all the local topic clusters $C_w$ at
%each window $w$, there is another question emerging immediately:
%do these topic clusters at different windows have connections with
%each other. Could we explore such connection to identify and track
%the evolution of on-going events. And most importantly, does this
%connection will be helpful us to identify the priming events which
%is formally defined as below:

Based on the definition of local topic cluster, we further define
a priming event as follows.

\begin{definition}
(\textbf{Priming Event}) A priming event $pe$ consists of three
components: (1) Two timestamps to denote the beginning and the
ending of the event in the window $W_{pe}$; (2) A sequence of
local topic clusters $c_{w,i}, w \in W_{pe}$; (3) a score
$Score(pe)$ representing its priming effect.
\end{definition}

% define

% overview of the approach
Our algorithm is designed to detect event with high priming event score and  there
are three major steps: (1) Data Transformation, (2) Global
Influential Topic Detection, (3) Local Topic Cluster Path Detection and 4) Event extraction by grouping similar topic cluster paths. Details are given in the following
sections.

\section{Data Transformation}

\label{sec:dataTransform}

In this section, we present how we transform and normalize the
features $F$ and the index $I$.

\subsection{Bursty Period Detection}
\label{sec:burstyFeatures} We first discuss how to select features
in window $w$ to represent on-going events. Given the whole
feature set $F$, we find that the emergence of an important event
is usually accompanied with a ``burst of words'': some words
suddenly appear frequently when an event emerges, and their
frequencies drop when the event fades away. Hence, by monitoring
the temporal changes of the word distributions in the news
articles, we can determine whether there is any new event
appearing. Specifically, if a feature suddenly appears frequently
in a window $w$, we say that an important event emerges.

In order to determine which of the intervals the feature $f$
``suddenly appears frequently'', we compute the ``probability of
bursty'' for each window, $w \in W$. Let $P(f, w; p_e)$ be the
probability that the feature $f$ is bursty in window $w$ according
to $p_e$, where $p_e$ is the probability that $f$ would appear in
any arbitrary windows given that it is not bursty. The intuition
is that if the frequency of $f$ appearing in $w$ is significantly
higher than the expected probability $p_e$, then $w$ is regarded
as a bursty period
\cite{DBLP:conf/wise/WuFYL08,DBLP:conf/kdd/FungYLY07}. We compute
$P(f, w; p_e)$ according to the cumulative distribution function
of the binomial distribution:
\begin{equation}
\label{eq: burst_probability1}
P(f, w; p_e) = {N_w \choose l}p_e^l (1 - p_e)^{N_w - l}
\end{equation}
where $N_w$ is the total number of words appearing in window $w$,
and $n_{f, w}$ is the frequency of $f$ appearing in $w$. $P(f, w;
p_e)$ is a cumulative distribution function of the binomial
distribution, and $P(l; N_w, p_e)$ is the corresponding
probability mass function. Therefore, $P(f, w; p_e)$ represents
the bursty rate of $f$ in window $w$. The bursty periods of $f$
are identified by setting a specific threshold such that $P(f, w;
p_e)$ is larger than the threshold. With the transformation, we
obtain the bursty probability time series for each feature $f \in
F$ as $p(f) = \{p(f, 1), p(f, 2), \ldots, p(f, |W|)\}$ and the
bursty windows of $f$ are denoted as $B_{f}$.

\subsection{Volatility Transformation and Discretization}
\label{sec:volatilityTransform} We now discuss how to monitor the
change of the index $I$ to reflect the effect of priming events.
In this paper, instead of looking at the rise/fall of $I$ solely,
we study the change of volatility of time series
\cite{DBLP:conf/icdm/RobertsonGW07,Gronke2002425}. Volatility is
the standard deviation of the continuously compounded returns of a
time series within a specific time horizon and is used to measure
how widely the time series values are dispersed from the average
as follows:
\begin{equation}
\label{eq:sdv} \sigma = \sqrt {\sum\limits_{i = 1}^n {[R_i - E(R_i
)]}^2 P_i}
\end{equation}
where $R_i$ is the possible rate of return, $E(R_i)$ is the
expected rate of return, and $P_i$ is the probability of $R_i$. We
can transform the index time series $I$ to the volatility of time
series, $VI=\{VI_1,...,VI_{|W|}\}$.
%\begin{figure}[t]
%\centering
%\includegraphics[height=8cm, width=8cm]{pic/logVol.pdf}
%\vspace*{-2cm}
% \caption{A logistical Distribution fit for Volatility Index Data}
%\label{fig:fitVol}
%\end{figure}

Given the volatility index $VI$, we observe  that there are some
abnormal behavior at certain time windows. For example, in the 911
event, there is a volatility burst for President Bush's approval
index. Such phenomena will bring tremendous bias to the events
happening in these volatility bursty windows. In order to avoid
such bias, we further transform the volatility index $VI$ to
obtain a discrete representation with equal probability
\cite{DBLP:conf/dmkd/LinKLC03}. According to our experiment, the
volatility index can fit into a logistic distribution. Therefore,
we can produce a equal-sized area under the logistic curve
\cite{D.Montgomery_book99} and transform the index volatility time
series into a probabilistic time series $PVI=\{PVI_1,...,PVI_{|W|}\}$.

\section{Global Influential Topic Detection}
\label{sec:topicDetection}

% Model

Given the feature set $F$ and the probability volatility index
$PVI$, our task here is to identify a set of influential topics
$\{\theta_1, ..., \theta_k\}$, where each topic $\theta_k$ is
formed by a set of keywords
$F_{\theta_k}={f_{\theta_k,1},...,f_{\theta_k,|\theta_k|}}$. The
problem can be solved by finding the optimal $\theta_k$ such that
the probability of the influential bursty features grouped
together is maximum for the text stream $\mathcal{D}$ and $PVI$.
Below, we formally define the probability of $\theta_k$:

\begin{definition}
(\textbf{Influential Topic Probability}) The probability of an
influential topic $\theta_k$ is given by
\begin{equation}
\label{eq:max_prob}
 P(\theta_k|\mathcal{D}, PVI)=\frac{P(PVI,\mathcal{D}|\theta_{k})P(\theta_k)}{P(\mathcal{D},
 PVI)}.
 %= \frac{P(PVI|\theta_k)P(D|\theta_k)P(\theta_k)}{P(D, PVI)},
\end{equation}
\end{definition}

Since $P(\mathcal{D}, PVI)$ is independent of $\theta_k$, we only
consider the numerator of Eq. (\ref{eq:max_prob}). We use the
topic $\theta_k$ to account for the dependency between
$\mathcal{D}$ and $PVI$. Therefore, given $\theta_k$,
$\mathcal{D}$ and $PVI$ are independent. Our objective function
then becomes:

\begin{equation}
\label{eq:log_prob}
 \max P(PVI,\mathcal{D}|\theta_k)P(\theta_k)= \max P(PVI|\theta_k)P(\mathcal{D}|\theta_k)P(\theta_k),
\end{equation}

Some observations can be made on Eq. (\ref{eq:log_prob}). The
second component $P(\mathcal{D}|\theta_k)$ represents the
probability that the influential topic generates the documents.
And intuitively, we expect the document overlap of the features
from the same topic to be high. The third component of
$P(\theta_k)$ represents the probability of the features to be
grouped together. And two features should be grouped together if
they usually co-occur temporally. Therefore, these two components
basically require the features of $\theta_k$ to be coherent at
both the document level and the temporal level. So generally, if
more features are grouped together, the values of the second and
the third components will decrease. And the first component
$P(PVI|\theta_k)$ represents the probability that the influential
topic triggers the volatility of the time series index. Obviously,
if the features in the group cover more windows with high
volatility probability, the value of the first component will be
higher. This will make the algorithm look for the features with
high potential impact on the index. Below, we show how we estimate
these three components.

First, we define the document similarity of two features using Jaccard Coefficient \cite{IntroDM_book00}.
\begin{equation}
\label{eq:feature_sim} sim(f_i,f_j)= \frac{|D_{f_i}\cap
D_{f_j}|}{|D_{f_i}\cup D_{f_j}|},
\end{equation}
where $D_{f_i}$ is the document set containing feature $f_i$.
Then the $P(\mathcal{D}|\theta_k)$ can be estimated as below:
\begin{equation}
\label{eq:topic_prob} P(\mathcal{D}|\theta_k)=
\frac{1}{|F_{\theta_k}|(|F_{\theta_k}|-1)} \sum_{f_i,f_j \in
F_{\theta_k}} sim(f_i,f_j).
\end{equation}

Second, in order to compute $P(\theta_k)$, we estimate the temporal similarity of two features by
comparing their co-occurrence over the whole set of windows $W$ as
below:

\begin{equation}
\label{eq:feature_prob} \rho(f_i,f_j)= \frac{|p(f_i)\cdot
p(f_j)|}{|p(f_i)||p(f_j)|},
\end{equation}
where $p(f_i)=\{p(f_i,1),p(f_i,2),...,p(f_i,|W|)\}$ is the bursty
probability time series of $f_i$ computed in Section
\ref{sec:burstyFeatures}. Then the probability of a set of
features belonging to $\theta_k$ can be estimated by the average
similarity for each pair of features:

\begin{equation}
\label{eq:topic_prob} P(\theta_k)=
\frac{1}{|F_{\theta_k}|(|F_{\theta_k}|-1)} \sum_{f_i,f_j \in
F_{\theta_k}} \rho(f_i,f_j).
\end{equation}

Finally, in order to estimate the $P(PVI|\theta_k)$, we define the influence probability for a feature $P(PVI|f)$
as:

\begin{equation}
\label{eq:PVI_feature_prob} P(PVI|f)=  \frac{\sum_{w \in B_{f}}
PVI_w}{\sum_{f\in F}\sum_{w \in B_{f}} PVI_w}.
\end{equation}
Since the denominator of Eq.(\ref{eq:PVI_feature_prob}) holds the same for all the features, we just take the numerator for computation. And $P(PVI|\theta_k)$ can be estimated as

\begin{equation}
\label{eq:PVI_topic_prob} P(PVI|\theta_k)= \sum_{f\in
F_{\theta_k}} P(PVI|f).
\end{equation}

Finally, the topic can be extracted using a greedy algorithm by
maximizing Eq. (\ref{eq:log_prob}) for each topic in a similar way
as in \cite{DBLP:conf/vldb/FungYYL05}.

% Approach Analysis

However, Eq. (\ref{eq:log_prob}) is different from the objective
function defined in \cite{DBLP:conf/vldb/FungYYL05} since we
extract topics with respect to an interested time series index $I$
rather than purely based on text documents. Consider the
\emph{worker} example again. The document similarity and the
temporal similarity of \emph{worker} and \emph{union} are 0.31 and
0.25, while those of \emph{worker} and \emph{wage} are 0.35 and
0.1. If we do not consider the index $I$ by setting
$P(PVI|\theta_k)=1$, by Eq. (\ref{eq:PVI_feature_prob}),
$P(worker,union|\mathcal{D},PVI)=0.31\times0.25=0.0775$ and
$P(worker,wage|\mathcal{D},PVI)=0.35\times0.1=0.035$. As a result,
the algorithm would combine \emph{worker} and \emph{union}.
However, since $P(PVI|union)=30$ and $P(PVI|wage)=74$, by
considering the feature influence to the index $I$, we have
$P(worker,union|\mathcal{D},PVI)=0.0775*30=2.325$ and
$P(worker,union|\mathcal{D},PVI)=0.035*74=2.59$. In this way, the
algorithm will instead group $worker$ and $wage$ together to make
an influential topic with respect to $I$.

As shown above, since influential topics carry the volatility index
information, it brings benefits to the priming event detection. In the
above example, if we detect a new event containing the common topic
\{\emph{worker, union}\}, the event may be trivial since the topic has
a low influence probability in the history. But if we detect an event
with \{\emph{worker, wage}\} instead, this event has a higher
probability to be a priming event. This is because the high influence
probability of the topic indicates that the events with such topic
attracted the public attention and changed people's mind in the past.

After extracting each $\theta_k$, the bursty rate of $\theta_k$ in a
window $w$ can be computed as below:

\begin{equation}
\label{eq:topic_bursty} P(\theta_k, w)=
\frac{1}{|F_{\theta_k}|}\sum_{f\in F_{\theta_k}} p(f, w)
\end{equation}

The bursty period of $\theta_k$ is determined in a similar way to
detecting the bursty period of features in Section
\ref{sec:burstyFeatures}. And we use $\theta_w$ to denote all the
bursty topics in window $w$.

\section{Micro Priming Event Detection}

\label{sec:eventExtraction}

%Let us assume that the influential topics $\theta$ related to
%index $I$ are identified. Let $D_w$ be a set of documents in $w$.

A priming event $pe$ consists of three components: (1) Two timestamps
to denote the beginning and the ending of the event in the window
$W_{pe}$; (2) A sequence of local correlated topic clusters $C_{w}, w
\in W_{pe}$; (3) A score representing its influence on the index which
is defined by considering the following three aspects of information:
\begin{enumerate}
\item {High event intensity $B_{pe}$}: The event contains high bursty
  topics. The intensity at each window $w$ is estimated as follows: 
\begin{equation}
\label{eq:event_bursty} B_{pe,w}=  \frac{1}{|\theta_k|}\sum_{\theta_k
  \in \theta_{pe,w}}P(\theta_k, w) 
\end{equation}
And the intensity over the whole period is measured by $\|B_{pe}\|$
where $\|$ means taking 2-norm on the vector.

\item {High index volatility}: The index must have high volatility
  during the bursty period of this event which is measured by
  $\|PVI_{pe}\|$.

\item {High index and event co-movement rate}: The index and event
  intensity time series should be highly correlated which is measured
  by their linear correlation $Corref(B_{pe}, PVI_{pe})$.
\end{enumerate}

We combine these three measures and define the priming event score as below:
\begin{equation}
\label{eq:pe_score} Score_(pe)= \|B_{pe}\|\cdot\|PVI_{pe}\|\cdot
Corref(B_{pe}, PVI_{pe}) 
\end{equation}

In this section, we proposed a two-phase algorithm to detect events
with high priming event score. We first look for potential important
topic cluster at each local time period by extracting and grouping
bursty topics. (High event intensity). We then take the topic cluster
at the period with most significant index volatility as seed and probe
the whole priming event path (High index volatility and High index and
event Co-movement rate). Below, we discuss the two phases of the
algorithm in detail.

\subsection{Local Topic Cluster Detection}
\label{sec:topicCluster}

As discussed in Section \ref{sec:overview}, we usually observe
multiple correlated bursty topics at a time window $w$
representing the same event. Therefore, we first group the
correlated topics into topic clusters at each time window $w$.

Intuitively, if two topics $\theta_i$ and $\theta_j$ belong to the
same event, the reporter usually discusses them in the same news
article and they would have a high degree of document overlap. We
first define the document frequency vector for a topic $\theta_k$
at a window $w$.

Let $\theta_w$ be a set of bursty topics in window $w$ and $D_w$
be the set of documents in window $w$. We define
$D_{f,w}=\{d_{f,w,1},d_{f,w,2},...,d_{f,w,|D_w|}\}$ be the term
frequency vector for feature $f$ in the documents in window $w$.

Then, the document frequency vector for a topic $\theta_k$ at a
window $w$ is computed by the average of the term frequency vector
as below:

\begin{equation}
\label{eq:topic_vector} D_{\theta_k,w}=
\frac{1}{|F_{\theta_k}|}\sum_{f \in F_{\theta_k}} D_{f,w}.
\end{equation}

Then the similarity between two topics $\theta_i$ and $\theta_j$
at window $w$ can be estimated by the cosine similarity:

\begin{equation}
\label{eq:topic_sim} sim(\theta_i,\theta_j, w) =
\cos(D_{\theta_i,w}, D_{\theta_j,w}).
\end{equation}

In order to cluster the set of topics $\theta_w$ into several
topic clusters, we use the K-Means clustering algorithm
\cite{IntroDM_book00}. We determine the optimal cluster number by
examining the quality of the clustering result under different
cluster number $k$. And the quality of the clustering is measured
based on the ratio of the weighted average inter-cluster to the
weighted average intra-cluster similarity
\cite{DBLP:conf/ideal/OzyerAB06}. After clustering, for a window
$w$, we obtain a set of topic clusters
$C_{w}=\{c_{w,1},c_{w,2},...\}$.

% why need to partition

% describe kmeans similarity

% describe kmeans number

% describe requirement

\begin{figure}[t]
\centering
\hspace*{-0.8cm}
\includegraphics[height=4cm, width=8cm]{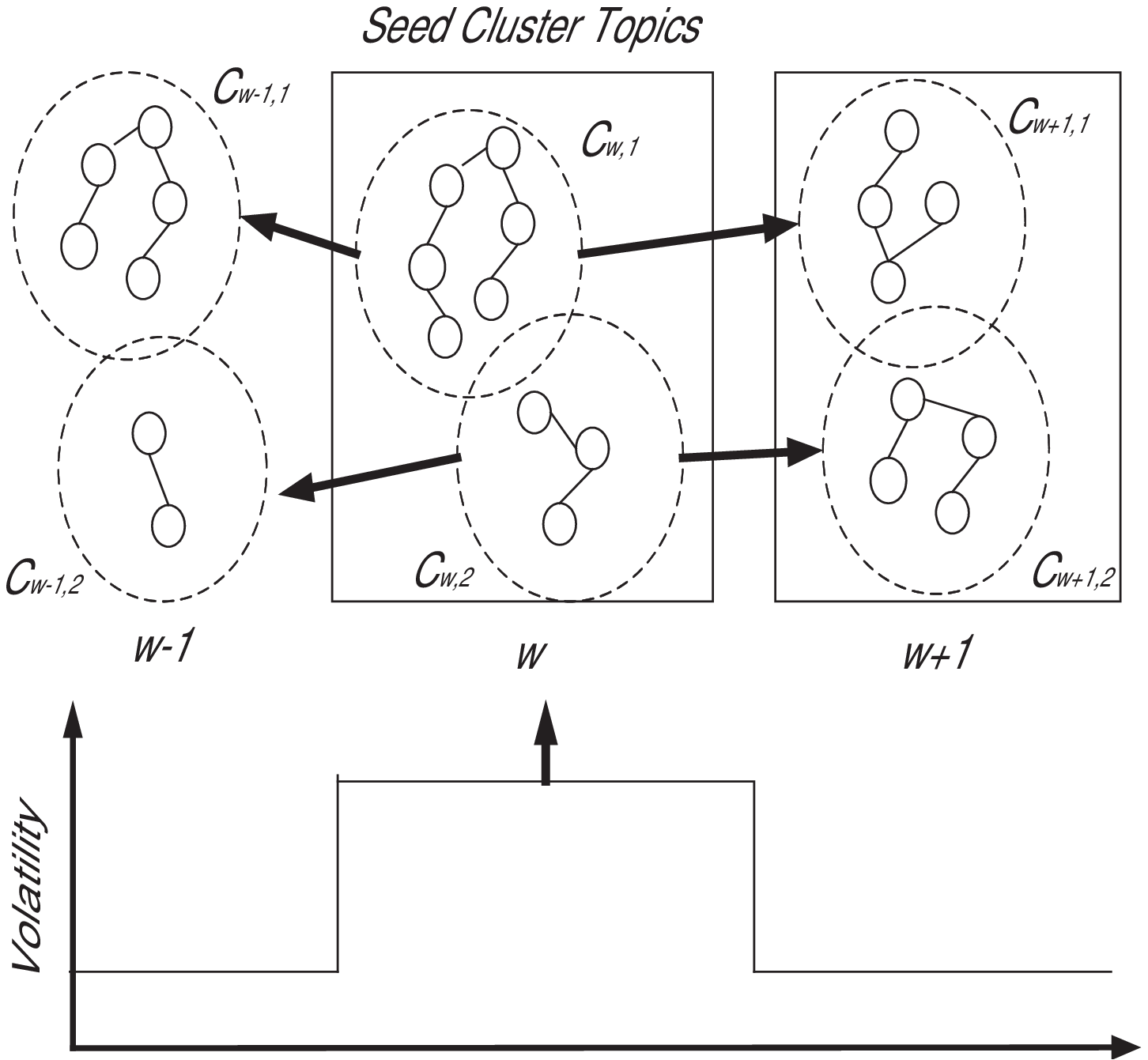}
 \caption{Topic Clusters Probe}
\vspace*{-0.6cm}
\label{fig:probe}
\end{figure}

\subsection{Composite Topic Cluster Path Detection}
\label{sec:topicClusterPath}

The detected topic cluster $c_{w,i}$ at each window represents a
snapshot of an event. And the remaining question is how could we
utilize these topics clusters and connect them into a topic cluster
path to represent a complete priming event.

Intuitively, the time period with high index volatility would have
better chance to contain the right priming event.  Therefore, we sort
the time periods according to the index volatility probability in
Section \ref{sec:dataTransform} and start from topic clusters in the
time period with the highest volatility probability.  Figure
\ref{fig:probe} illustrates the idea where the bottom part is the
volatility index and the upper part is the detected topic clusters at
three consecutive windows.  As we can see, window $w$ has the highest
volatility and therefore the two topic clusters in $w$ will be taken
as seed topic clusters.  Then, we will start to construct the priming
event from these two topic clusters. Specifically, we will probe
forward/backward and associate similar and appropriate topic clusters
into a topic cluster path $P$.

Here, we measure the similarity between two topic clusters in two
consecutive windows, $c_{w,i}$ and $c_{w-1,j}$ by looking at their
intersection of influential topics. More specifically, the similarity
of two topic clusters can be computed by adding the topic probability
$P(\theta_k|\mathcal{D},PVI)$ as a weight to the Jaccard Coefficient.

\begin{equation}
\label{eq:cluster_sim} sim(c_{w,i},c_{w-1,j})=
\frac{\sum_{\theta_k \in c_{w,i}\cap
c_{w-1,j}}P(\theta_k|\mathcal{D},PVI)}{\sum_{\theta_k \in
c_{w,i}\cup c_{w-1,j}}P(\theta_k|\mathcal{D},PVI)}.
\end{equation}

Eq. (\ref{eq:cluster_sim}) assigns a higher similarity score to
two topic clusters whose overlapping topics have higher topic
probability, i.e., the topics are more coherent and influential.

However, even if two cluster have a high similarity score, we still
can not link them directly since the topic cluster path with the new
topic may have worse quality compared to the origin one.  Here, we
measure the quality of a topic cluster path using the same measure as
for the priming event, which is defined in Eq. \ref{eq:pe_score}. For
example, in Figure \ref{fig:probe}, the link between $C_{w,1}$ and
$C_{w+1,1}$ means that they have a high similarity with each
other. however, we found if we extend the path to $w+1$ by connecting
$C_{w,1}$ with $C_{w+1,1}$, then event intensity is still very high at
window $w+1$ but the index volatility has decreased. This results
lower correlation value between the topic cluster path and volatility
index and eventually reduces the priming event score. Therefore, we
will only connect two topic clusters if 1) they have a similarity
score $sim(c_{w,i},c_{w+1,j})$ higher than a predefined threshold
$\sigma$. 2) the topic cluster path score improves by integrating the
new topic cluster.

\comment{
\begin{figure}[t]
\centering
\hspace*{-0.8cm}
\includegraphics[height=4cm, width=8cm]{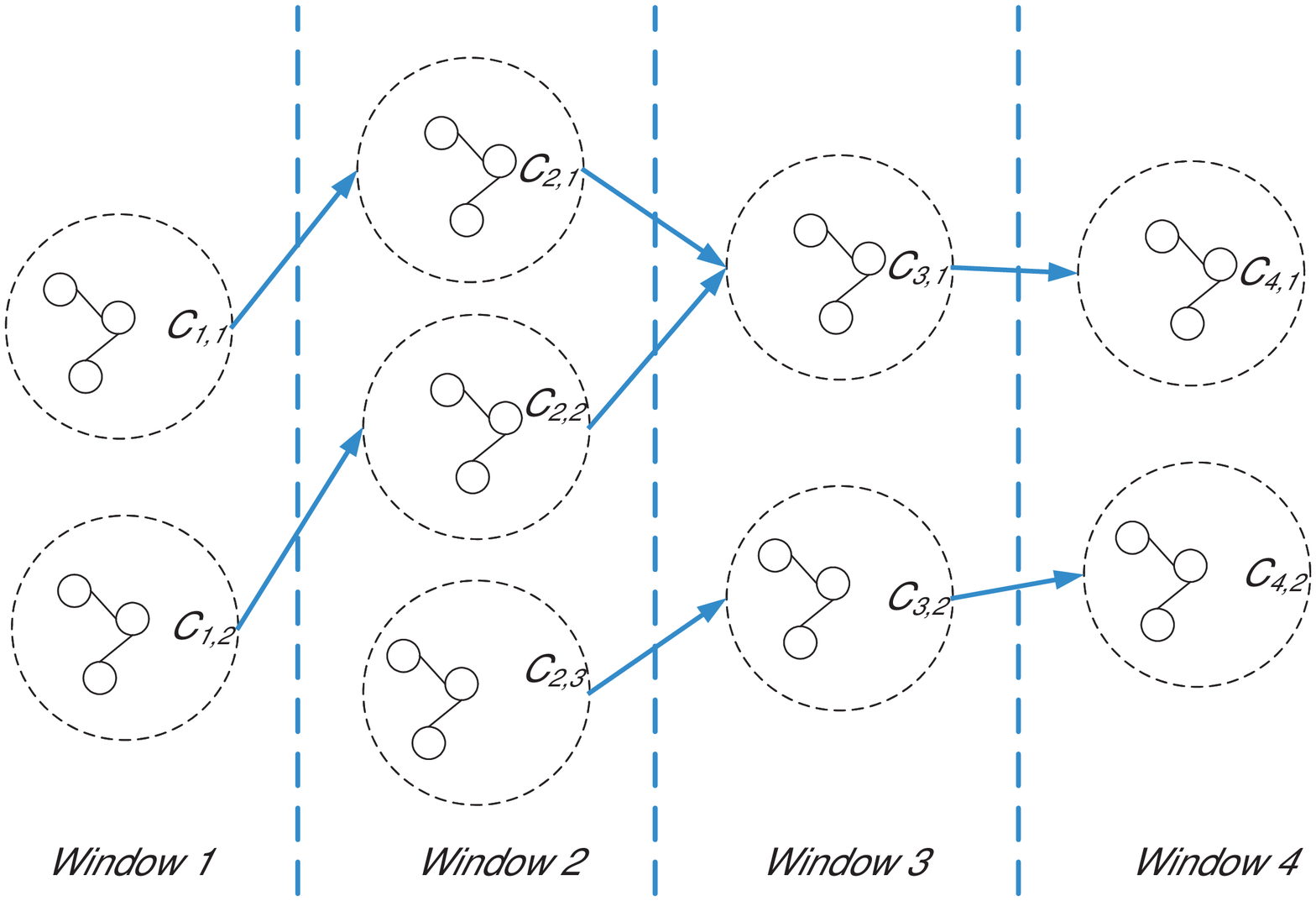}
 \caption{Evolving Topic Clusters}
 \vspace*{-0.6cm}
\label{fig:evolving}
\end{figure}
}
After connecting appropriate topic clusters, the algorithm form a directed acyclic graph
(DAG) of topic clusters between windows and we formally define a Path $P$ in this
graph as follows.
 \begin{definition}
(\textbf{Topic Cluster Path})   A topic cluster path $P$ of length
$l$ in a topic cluster graph is a sequence of $l$ clusters:
$c_{w_1, i_1} \rightarrow c_{w_2, i_2} \rightarrow \ldots
\rightarrow c_{w_l, i_l}$, such that $\{w_1, \ldots, w_l\}$ are
$l$ consecutive windows and there is an edge between two
consecutive clusters in the graph.
\end{definition}

In addition, the topic cluster path may have overlap between each other which
they may express
different aspects of the same priming event. For example, in the
gulf war event, one topic cluster path may show the progress of
the battle in Iraq, while another path may record the actions from
US's allay. Therefore, we measure the similarity between two
overlapping paths as follows:

\begin{equation}
\label{eq:path_sim} sim(P_{i},P_{j})= \frac{|P_{i}\cap
P_{j}|}{\min(|P_{i}|,|P_j|)}.
\end{equation}

If the similarity between two paths $sim(P_{i},P_{j})$ is high, then we group these two paths and form
a priming event. Here we explore the agglomerative hierarchical clustering \cite{IntroDM_book00} to conduct  grouping and 
extract events.

\begin{algorithm}[t] \caption{DiscoverPrimingEvents}
 \label{algorithm:DiscoverPrimingEvents}
{\sc
Input:} News document stream $D$ and index $I$\\
 {\sc Output:} priming events $PE$
\begin{algorithmic}[1]
    \FOR {every $w$}
        \STATE {retrieve the burst topics at $w$, $\theta_w$}
        \STATE {$C_w\leftarrow KMeans(\theta_w)$}
        \STATE {$Sign_{w,i}=true$}
    \ENDFOR
    \FOR {every $w$ in $PVI_w$ descending order}
        \STATE {$IP=\{\}$}
        \FOR{each $c_{w,i}\in C_{w}$ and $Sign_{w,i}$}
            \STATE {generate a new Path $P_k$}
            \STATE {add $P_k$ to inverted path list $IP_{c_{w,i}}$}
            \STATE {$Sign_{w,i}=false$}        
        \ENDFOR
        \STATE {$P=ProbeEventPath(C,P,IP,w,Sign)$}
    \ENDFOR
    \STATE{compute $sim(P_i,P_j)$ for every $P_i,P_j\in P$ according to Eq.\ref{eq:path_sim}}
    \STATE{return $PE=Cluster(sim)$}
\end{algorithmic}
\end{algorithm}

\begin{algorithm}[t] \caption{ProbeEventPath}
 \label{algorithm:ProbeEvent}
{\sc
Input:} Topic Cluster $C$, Topic Cluster Path $P$, Seed Inverted Path List $SIP$, seed window $sw$, $Sign$ \\
 {\sc Output:} updated Topic Cluster Path $P$
\begin{algorithmic}[1]
\STATE {$w=sw$ and $IP=SIP$}  
\ \  \ \ \ \ \ \ \ \ \ \ // + for right side
\WHILE {$IP$ is not empty} 
    \FOR {each $c_{w+1,j}\in C_{w+1}$ and $Sign_{w+1,j}$}  
        \FOR {every $c_{w,i}$ has an entry in $IP$}
            \STATE {Compute topic similarity $Sim(c_{w,i},c_{w+1,j})$ according to Eq. \ref{eq:cluster_sim}}
            \IF {$sim(c_{w,i},c_{w+1,j})>\sigma$}
                \FOR {each path $P_k$ in the inverted path list $IP_{c_{w,i}}$}
                    \STATE {compute $score(P_k, c_{w+1,j})$ according to Eq. \ref{eq:cluster_sim}}
                    \IF {$score(P_k, c_{w+1,j})>P_k.score$}
                        \STATE {$P_k.score=score(P_k, c_{w+1,j})$}
                        \STATE {Add $c_{w+1,j}$ to $P_k$}
                        \STATE {Add $P_k$ to new inverted path list of $c_{w+1,j}$, $NIP_{c_{w+1,j}}$ }
                        \STATE {$Sign_{w+1,j}=false$}
                    \ENDIF
                \ENDFOR
            \ENDIF
        \ENDFOR
    \ENDFOR
    \STATE {$IP=NIP$}
    \STATE {$w=w+1$}    
\ENDWHILE
\STATE {$w=sw$ and $IP=SIP$}
\ \  \ \ \ \ \ \ \ \ \ \ // + for left side
\STATE {...}
\end{algorithmic}
\end{algorithm}

Algorithm \ref{algorithm:DiscoverPrimingEvents} describes the whole
process of how we detect priming events.  First of all, for each time
window $w$, we retrieve the bursty influential topics $\theta_w$ in
Line 2.  Then Line 3 groups the topic into topic clusters $C_w$ as
described in Section \ref{sec:topicCluster}. Here we keep a $Sign$ to
indicate whether the topic has been used to construct priming event.
And Line 6-14 is the key process to construct the priming events by
connecting the topic clusters in consecutive windows. As discussed
before, it will start from the window with highest index
volatility. In Line 8-12, for each seed window $w$, we check the
unused topic cluster and generate a new path for each of them.  Here,
we also maintain the inverted path list for $c_{w,i}$,
$IP_{c_{w,i}}$. With this preparation, in Line 13, we probe event path
which is described in detail in Algorithm
\ref{algorithm:ProbeEvent}. After we discover all the paths, we
compute the topic cluster path similarity by Eq. \ref{eq:path_sim} in
Line 15 and group them together according to hierarchy clustering
algorithm in Line 16.

Algorithm \ref{algorithm:ProbeEvent} shows how to probe the whole
event path starting from a seed window $sw$. The algorithm is
conducted both forward and backward. Let's take the forward situation
as an example where we try to see whether we can extend a path to
window $w+1$.  And for each cluster in $C_{w+1}$, $c_{w+1,j}$, we
compare $c_{w+1,j}$ with all the topic cluster with an entry in $IP$
(i.e., it is a path to be extended) by computing cluster similarity
according to Eq. (\ref{eq:cluster_sim}) in Line 5.  If their
similarity is higher than $\sigma$, then in Lines 7-15, we check all
the paths in the path inverted list of $c_{w,i}$ and decide whether to
extend the path by checking whether the priming event score would
increases by integrating $c_{w+1,j}$ into each path.  Besides, we also
maintain the new inverted path list of $c_{w+1,j}$, $NIP_{c_{w+1,j}}$.
Finally, in Lines 19-20, we use the new inverted list $NIP$ to replace
the previous one $IP$ and move the window forward.  The same will
apply for the back-ward topic cluster path probing operation. The
algorithm will return a list of update topic cluster path $P$.

\section{Experimental Results}
\label{sec:experiment}

In this section, we evaluate our algorithm on real-world dataset from
political and finance domain.

We archive the news
articles from the ProQuest
database\footnote{\url{http://www.proquest.com/}}. In ProQuest, for political news, we take ``President Bush'' and ``President Obama'' as
the query keywords and extract $15,542$ and $1,643$ news articles respectively from Jan.\ 1,
2001 to June.\ 1, 2010. For the financial news, we want to study the priming events during the financial crisis period, so we
take ``Finance'' as
the query keywords and extract $9,416$ news articles from June.\ 1, 2007 to 
June.\ 30, 2010.

For the preprocessing of these news articles, all
features are stemmed using the Porter stemmer. Features are words
that appear in the news articles, with the exception of digits,
web page address, email address and stop words.  Features that
appear more than 80\% of the total news articles in a day are
categorized as stop words. Features that appear less than 5\% of
the total news articles in a day are categorized as noisy
features.  Both the stop words and noisy features are removed.

For the time series index, we archive the President Approval rating index from the Gallup
Poll\footnote{\url{http://www.gallup.com}}
And the poll is taken every 10
days approximately. And for President Obama, we take a weekly
average rating based on the Gallup daily tracking. For the financial index, we use the popular U.S. market index, the S\&P 500 index and also take a weekly 
sample based on its daily index.
We further
identify the bursty probability time series and volatility time
series according to the methods introduced in Section
\ref{sec:burstyFeatures} and Section
\ref{sec:volatilityTransform}.

After these preprocessing, we have three real datasets.

\vspace{-2mm}

\begin{itemize}
\item{\textbf{Bush}}. It contains 1186 bursty feature probability
streams  and 1 volatility time series with equal length of 281 in
the 8 years of Bush's administration.

%%The sliding window vary from 120 to 1440, i.e., 10 minutes to two
%%hours.

\item\textbf{{Obama}}. It contains 1197  bursty feature
probability streams and 1 volatility time series with equal length
of 56 in the 13 months of Obama's administration.

\item\textbf{{Tsunami}}. It contains 961  bursty feature
probability streams and 1 volatility time series with equal length
of 156 in the 3 years of financial crisis.

\end{itemize}

We implemented our framework using C\# and performed the
experiments on a PC with a Pentium IV 3.4GHz CPU and 3GB RAM.

\subsection{Identifying Influential Topics}

\begin{table}
\centering
\begin{tabular}{|l|l|l|}\hline
 Bush & Obama & Finance
\\\hline\hline
     bin laden &   oil Mexico   & Moody Triple mature  \\
     north Korean &  civil movement equal  & Merrill lynch  \\
     Israel palestinian &   immigration illegal  & legman brother \\
     immigration illegal &  police arrest &  barrack obama primary \\
     destruct mass &  small business lend &  Germany \\
     \hline
\end{tabular}
\caption{Influential Topics} \label{table:topics}
\vspace*{-0.6cm}
\end{table}

With the algorithm in Section \ref{sec:topicDetection}, for political domain, we
identify 385 and 207 influential topics from Bush and Obama,
respectively. And for finance domain, we identify 100 influential topics from Tsunami. Table \ref{table:topics} gives the top 5 influential
topics and the rank is based on the influential topic probability
in Eq. (\ref{eq:max_prob}). As shown in the second column of the
table, \{bin laden\},  \{north Korea\} are 2-gram names which are
regarded as the largest threaten for Bush's administration from
2001-2008. Each of the other three topics consists of a set of
features which are not 2-gram names but coherent and influential
keywords in the document stream. The third column of the table
shows the influential topics from Obama. Compared with Bush's
where 4 out of 5  topics are about international affairs, there
are significant evolution of the influential topics of Obama's
topics. In particular, only first topic of Obama is about the
international issue, i.e., the oil slick in Mexico Gulf . Others are
all about how he deals with domestic affairs. The fourth column shows the influential topics
we detect from Tsunami. The first topic \{Moody Triple \} shows the common triggering reason for the market crash, i.e., when the Moody changes the Triple for the 
big and matured financial institutions, the investors got panic and cleaned their position quickly.  The second and third topics shows two big investment banks where the
Legman Brother went bankruptcy while Merrill Lynch was acquired by Bank of America after losing a lot in the financial crisis. Their destinies are watched by all the investors in the market and therefore made into influential topics. The forth and fifth topics are two parties whose actions are critical to the direction of the market. 
The first one is President Obama who leads the policy of United States and the second one is Germany which influence the decision on bailout plan of Europe. 

As discussed before, these
influential topics detected at a global level give us evidences in
detecting priming events at a micro level.

\subsection{Priming Events in Politics}
\begin{figure}
\centering
\hspace*{-0.8cm}
\includegraphics[width=10cm,height=6cm]{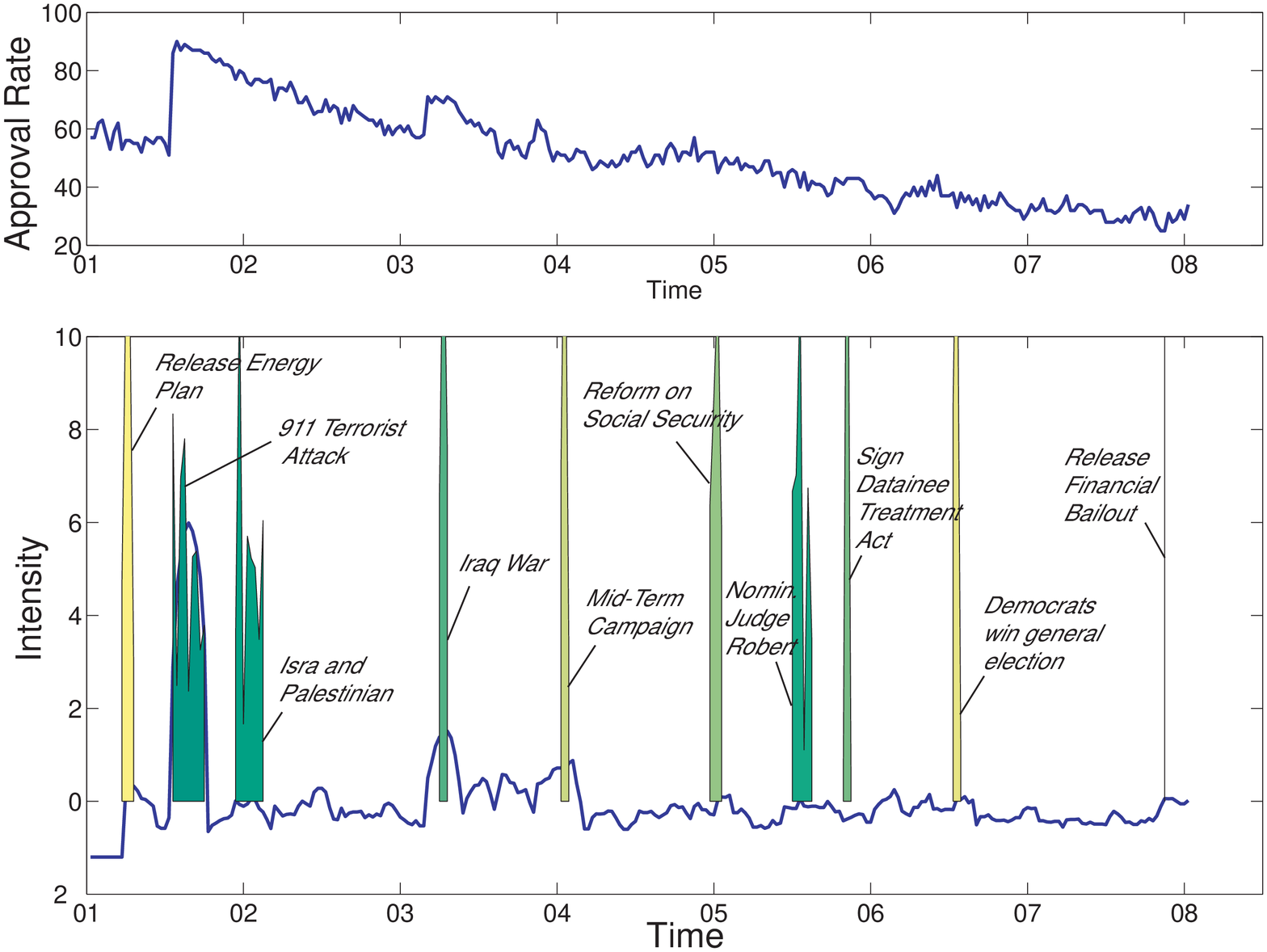}
\vspace*{-0.6cm}
\caption{Top 10 Priming Events of President Bush}
\label{fig:bush}
\end{figure}

\begin{figure}
\centering
\hspace*{-0.8cm}
\includegraphics[width=8cm,height=2.5cm]{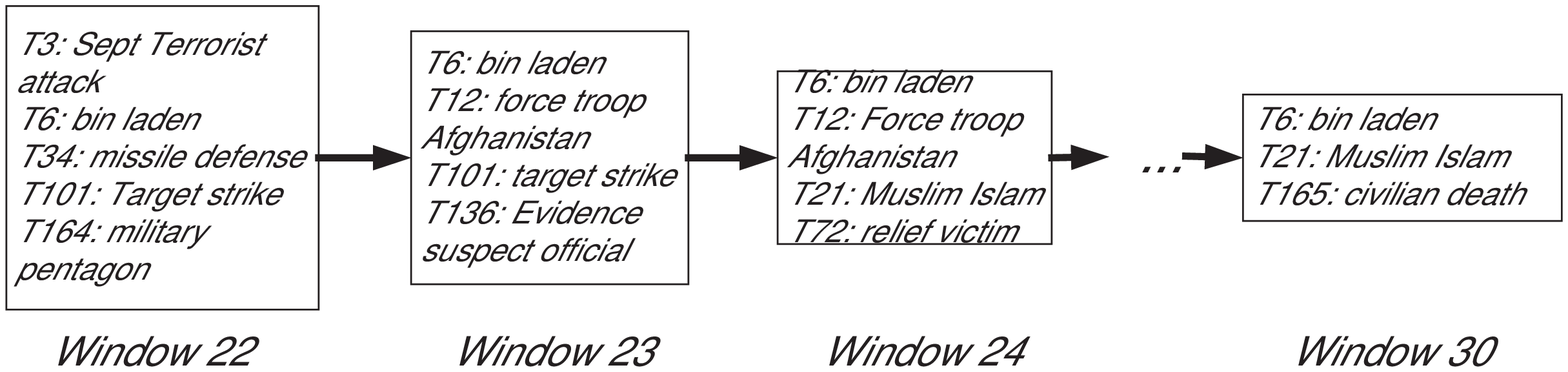}
\caption{Priming Event: 911 Terrorist Attack}
\vspace*{-0.6cm}
\label{fig:911}
\end{figure}

With the influential topics, we can identify the priming events
with by the algorithm in Section \ref{sec:eventExtraction}. Figure
\ref{fig:bush} shows the approval index of President Bush and the
top 10 priming events automatically detected over his eight years
administration. As shown in the upper part of Figure
\ref{fig:bush}, the approval index starts from 57\% and has a big
jump up to 90\% in Sep 2001. Then it drops quickly until a rebound
back to 70\% in 2003. After that, it continues dropping with some
small rebound in the middle and eventually reaches 34\% in 2008.
In the lower part, the blue line shows the volatility index
according to  Section \ref{sec:volatilityTransform} and the
colored waves represent the ranked priming events (we normalize
the value of these two time series and plot them in same graph).
The rank is based on the score given by Eq. (\ref{eq:pe_score}) and
the wave with deeper color represents the priming event with a
higher rank. The magnitude of the wave represents the intensity of
the event according to Eq. (\ref{eq:event_bursty}). From the
figure, we have two obvious observations: 1) the value of the
volatility index increased when a significant trend change
happened for the approval index. 2) During the periods when the
volatility index increased significantly, we detect a burst of
priming events. For example, after September 2001, we observe a
significant increase of the volatility index reflecting the big
jump of the approval index. At the same time, we detect the top 1
priming event about the 911 terrorist attack. Figure \ref{fig:911}
further shows its structure which contains a composite topic
cluster path with a length of 9 starting from window 22 to window
30. In each window, the path contains a topic cluster with a set
of topics lying in a highly overlapped document set. For example,
in window 22, the cluster consists of 5 bursty topics including
the influential topics of \{bin laden\} and \{Sept Terrorist
Attack\} that indicate the start of the 911 event. After that, in
window 23, we observe another cluster containing \{bin laden\} and
we connect it with the previous one since \{bin laden\} has a high
influential topic probability and dominates the topic similarity
measure according to Eq. (\ref{eq:cluster_sim}). We can also see a
certain degree of evolution between these two topic clusters since
the second cluster contains a new topic \{force troop
Afghanistan\} that is about U.S. sending troop to start the war.
In the following windows, we can see the topic clusters evolute
but all contain the influential topic \{bin laden\}. This event
ends in window 30 with a topic \{civilian death\} indicating that
the war results in the civilian death of the country. From this
priming event, we can see that the attack makes the U.S. people
united and support their President. Similarly, the volatility
index also reflects the rebounding of the approval index in 2003.
And the priming event covering that period is about the Iraq war
with the ending of U.S. victory which increased the public support
of President Bush. Other periods with significant increasing
volatility such as those in Mar 2001 and Jul 2004 can also be
explained by the detected priming events, i.e., President Bush
released the energy plan and won the mid-term campaign over John
Kerry.

\begin{figure}
\centering
\hspace*{-0.8cm}
\includegraphics[width=10cm,height=6cm]{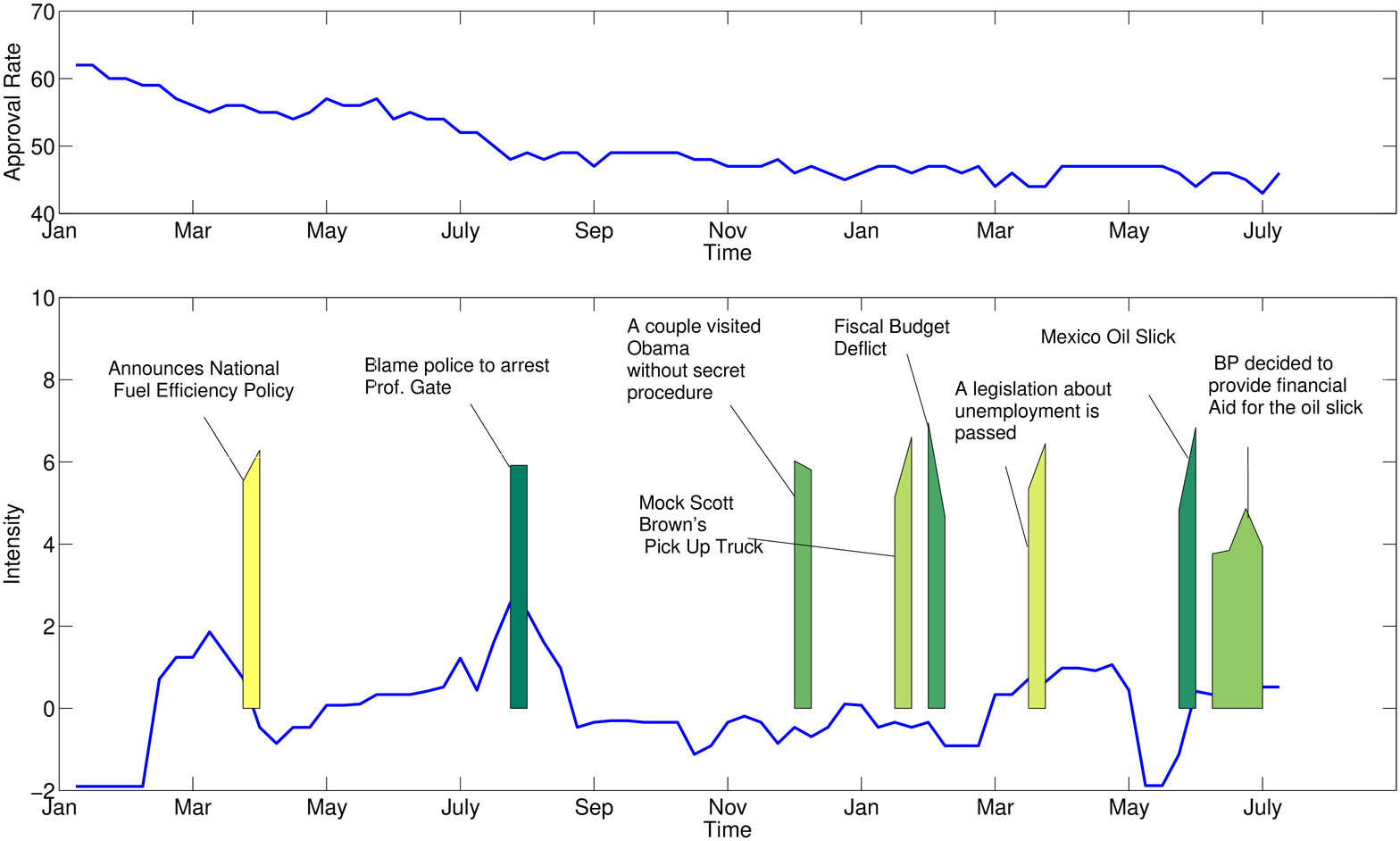}
\caption{8 Priming Events of President Obama (2009.1-2010.6)}
\label{fig:obama}
\end{figure}

Figure \ref{fig:obama} shows the approval index of President Obama
and the  8 priming events detected over his 16 months
administration. As discussed in Section \ref{sec:introduction},
users may be curious about what happened in the quick drop
periods in July, 2009. In the lower part of Figure
\ref{fig:obama}, when the volatility increased,
we detect the priming event that he criticized  on the police who
arrested Prof. Gate.  On the other hand, the algorithm also
detects that his effort to pass the legislation on unemployment in April, 2010 helped to improve his approval rate.

\comment{
In addition, among the top 10 priming events, we observe that two
of them both contain the topic \{Oil Mexico\}. Figure
\ref{fig:mexico} shows the event structure of them. As we
can see, the first priming event is about emergence of the oil spill crisis with a composite topic cluster
path of length 2. In addition to the topic \{Oil Mexico\},
it also contains a more specific topic \{guard conduct report\},
indicating the detail report of crisis. During this event, President Obama is blamed to react slowly resulting a drop for approval index. The
second event happened after two weeks and has a composite topic cluster
path of length 3. In addition to the topic \{Oil Mexico\},
it also contains a topic \{repeat prevent pressure container damage\},
indicating that it is about President Obama keeping giving pressure to British Petrol and requiring them to 
pay for the cost. This effort was approved by the public and produced 
a rise trend in the approval index. 

\begin{figure}
\centering
\hspace*{-0.8cm}
\includegraphics[width=8cm,height=2cm]{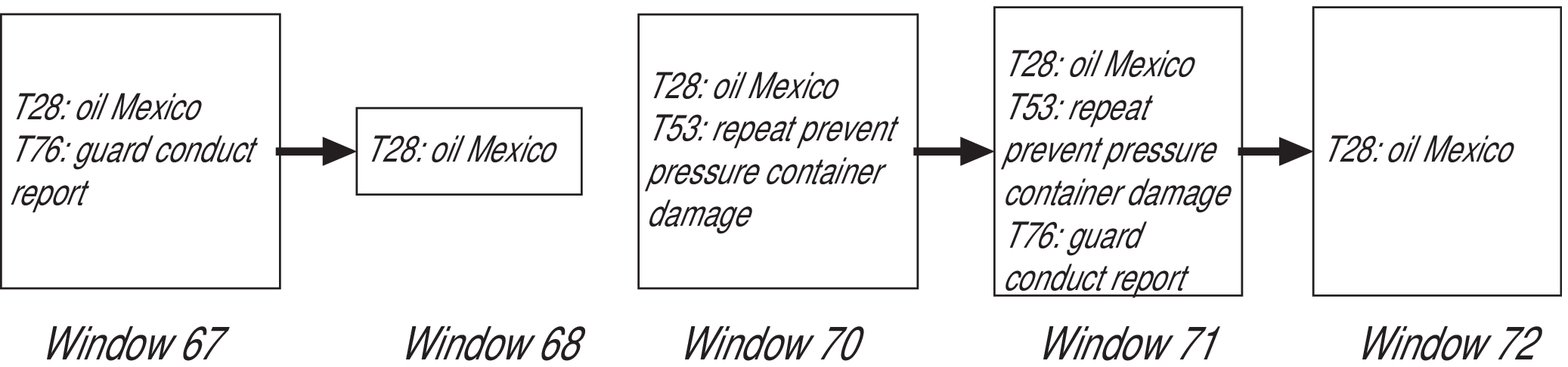}
\caption{Two Priming Events of Mexico Oil Spill}
\hspace*{-0.6cm}
\label{fig:mexico}
\end{figure}
}

\subsection{Priming Events in Finance}
\begin{figure}
\centering
\hspace*{-0.8cm}
\includegraphics[width=10cm,height=6cm]{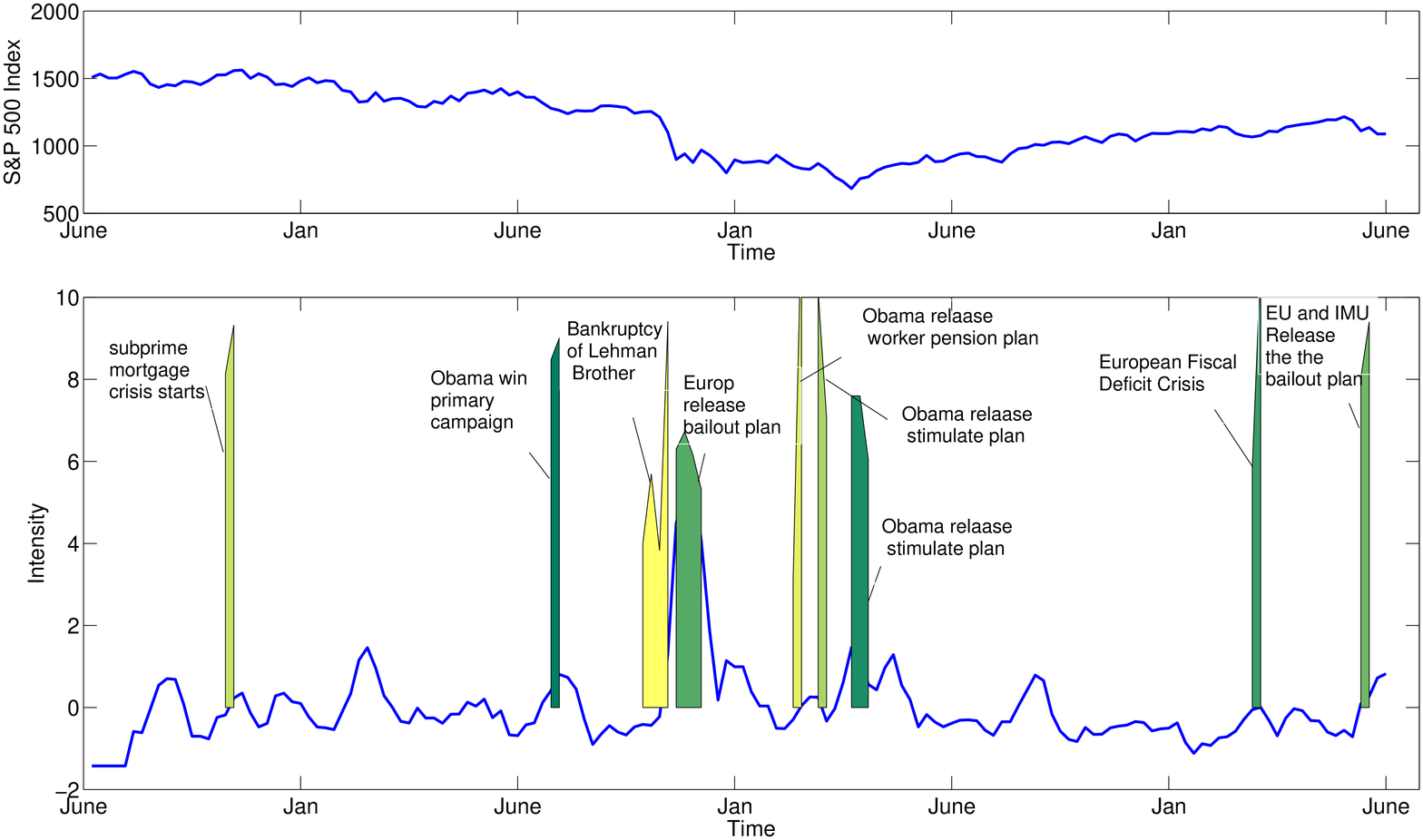}
\vspace*{-0.6cm}
\caption{Top 10 Priming Events of Financial Tsunami (2007.6-2010.6)}
\label{fig:finance}
\end{figure}

Figure \ref{fig:finance} shows the S\&P 500 index
and the top 10 priming events detected over the three years of financial crisis period.

\begin{figure}
\centering
\hspace*{-0.8cm}
\includegraphics[width=8cm,height=2.5cm]{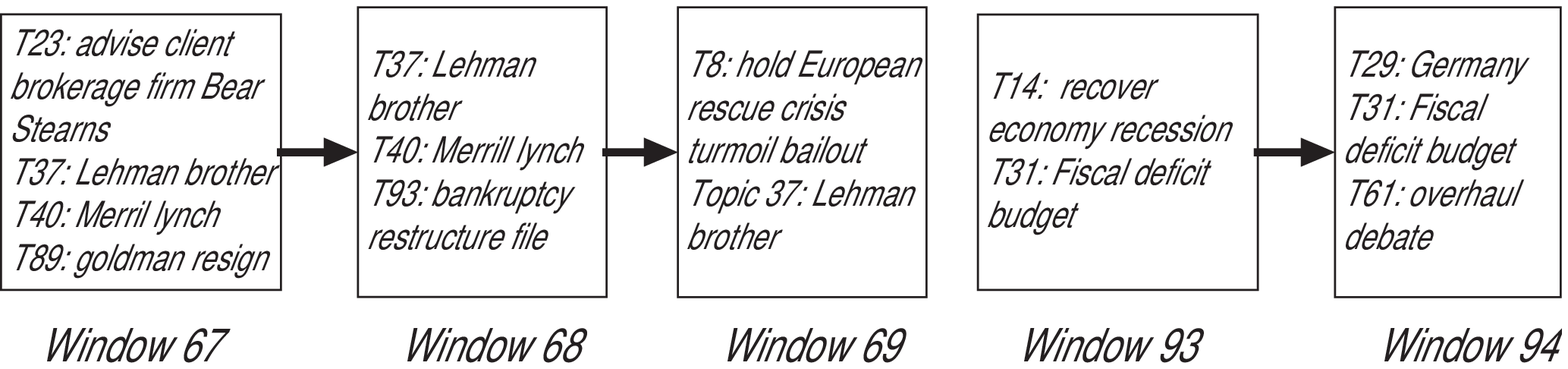}
\caption{Priming Event: Financial Crisis}
\hspace*{-1cm}
\label{fig:finance_event}
\end{figure}

In the upper part of Figure \ref{fig:finance}, we can see the S\&P 500 index reached the top of 1562 in Oct., 2007 and started to turn down.
At the same time, In the lower part, the volatility increased
and we detect the event that subprime mortgage crisis 
started to emerge.  As the crisis developing, the biggest index
 drop of 300 points is achieved in Oct., 2008 before which we detected the bankruptcy event of the 
legman brother. The left part of Figure \ref{fig:finance_event} shows its structure. In its three evolving windows, the influential topic \{Lehman Brother\} always exists.  And at the first window, it mentioned its possibility to fall down as Bear Stearns. In the following two windows, it integrated the bankruptcy topic as well as a topic about reaction from Europe.  After this event, the index kept dropping and achieved the bottom of 683
 in March, 2009. And it started to rebound with the detected event that
President Obama released the stimulation plan to re-build the confidence of investors. 
Recently, we also observed two priming events which are about the European Fiscal Deficit Crisis. The right part of Figure \ref{fig:finance_event} shows its structure which has a length of 2. At the first window, we detected the event of European Fiscal Deficit and at the second window we integrate a new topic cluster of \{Germany\} indicating that Germany played a major role in saving the crisis.

\subsection{Event Quality Comparison}

\begin{figure}
\centering
\hspace*{-0.8cm}
\includegraphics[width=8cm,height=3cm]{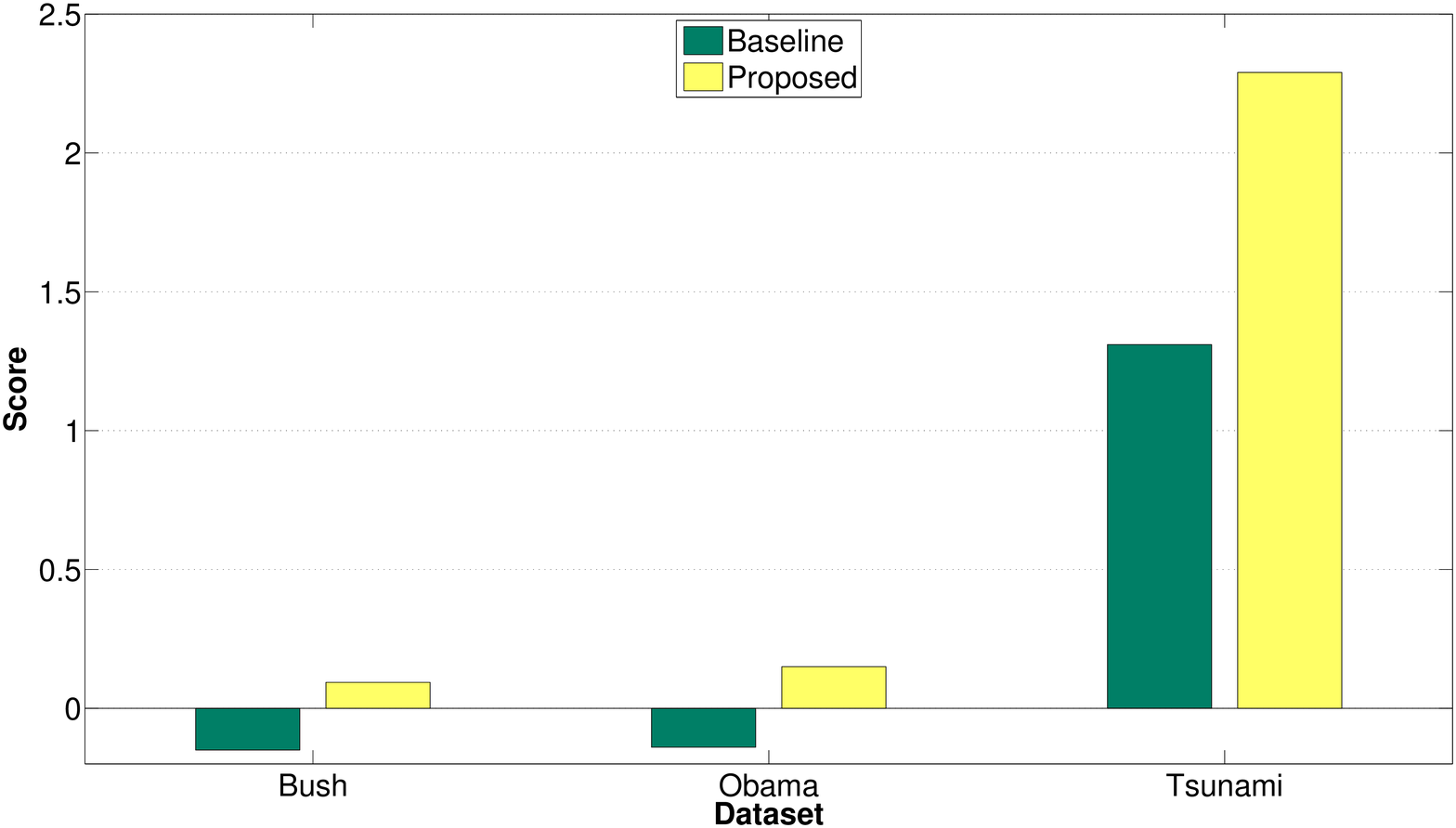}
\caption{Event Quality Comparison}
\vspace*{-0.6cm}
\label{fig:comparison}
\end{figure}

In order to further evaluate the performance, we implement a baseline approach as comparison which is based on 
the bursty events directly \cite{DBLP:conf/vldb/FungYYL05}. According the approach, the bursty period of the event is 
decided by:

\begin{equation}
\label{eq:topic_bursty} P(w, E_k)=
\frac{1}{|E_k|}\sum_{f\in E_k} p(f, w)
\end{equation}

where $E_k$ is similar to the influential topic identified in this paper but it do not consider integrating the index volatility into the detection process. Given events detected by the baseline approach and the proposed approach, we compared their average priming event scores.
Figure \ref{fig:comparison} shows the result and  our proposed approach  outperforms the baseline approach for the three datasets. And we also note that for the Bush and Obama dataset, the baseline approach shows negative value which means the detected event can not match well with the volatility index and therefore produce a lot of events with negative priming event score. Besides, we can see the average score of events on Tsunami dataset is much larger than the average score on Bush and Obama dataset. This is because the  volatility of financial index is much larger than the volatility of President approval index which leads to a larger priming score according to Eq. \ref{eq:pe_score}.

\section{Related Work}
\label{sec:relatedWork}

The problem of Topic Detection and Tracking (TDT)
\cite{topic.detection} is a classical research problem for many
years. The first stream of work used graphical probabilistic models to
capture the generation process of document stream
\cite{DBLP:journals/jmlr/BleiNJ03,DBLP:conf/nips/GriffithsSBT04}. \cite{DBLP:conf/kdd/WangM06}
extended the LDA model and incorporated location
information. \cite{DBLP:conf/kdd/WangZHS07} analyzed multiple
coordinated text streams and detected correlated bursty
topics. \cite{DBLP:conf/www/MeiCZZ08} added the social network
information as a regulation into the topic detection framework. On the
other hand, there are another stream of work which detected topic
based on the bursty features
\cite{DBLP:conf/kdd/Kleinberg02}. \cite{DBLP:conf/vldb/FungYYL05}
detected bursty features and clustered them into bursty
events. \cite{DBLP:conf/kdd/FungYLY07} further built an event
hierarchy based on the bursty features. \cite{DBLP:conf/sigir/HeCL07}
analyzed the characteristics of bursty features (power and
periodicity) and detected various types of events based on
it. \cite{DBLP:conf/kdd/LeskovecBK09} proposed an algorithm to track
short phrased and organized them into different news threads.
Although these work can detect topics and track event
efficiently. They can not tell the users which topics/events are
changing the real world's interested time series, President Approval
Rating, Stock Market Index.

In contrast, there are some work studying the relationship between the
text stream and interested time
series. \cite{CambridgeJournals:5827084} studied the relationship
between various topics over the volatility of presidential approval
index. \cite{DBLP:conf/kdd/GruhlGKNT05} attempted to find out the
relationship between the online query and its related sales
rank. \cite{G.Fung_IEEEIIB05} proposed a model for mining the impact
of news stories on the stock prices, by using a $t$-test based split
and merge segmentation algorithm for time series preprocessing and SVM
for impact classification. \cite{DBLP:conf/icdm/RobertsonGW07} studied
the relationship between the news arrival and the change of volatility
of stock market index.  \cite{stockriskrank} further explored this
relationship to rank the risk of stocks.  However, these work are
either relying on analysts to extract topics manually or just
identifying a list of influential keywords. In our work, we attempt
automatically identify these events by first organizing these words
into high level influential topics and then detect priming events
based on them.

\section{Conclusion}
\label{sec:conclusion}

In this paper, we study the problem of detecting priming events
based on a time series index and an evolving document stream. We
measure the effect of the priming events by the volatility rate of
the time series index. We propose a two-step framework to detect
the priming events by first detecting influential topics at a
global level and then forming priming events using detected
influential topics at a micro level. The experimental results on
the political and finance domains show that our algorithm is able
to detect the priming events that trigger the movement
of the time series index effectively. 

 {
\bibliographystyle{abbrv}
\bibliography{bib/conference,bib/conference2,bib/journal,bib/workshop,bib/symposium,bib/paper,bib/book,bib/workshop}
}

\end{document}